\documentclass{article}

\usepackage{PRIMEarxiv}
\usepackage{booktabs}       %
\usepackage{tabularx} %
\usepackage{float}
\usepackage[utf8]{inputenc} %
\usepackage[T1]{fontenc}    %
\usepackage{hyperref}       %
\usepackage{url}            %
\usepackage{amsfonts}       %
\usepackage{nicefrac}       %
\usepackage{microtype}      %
\usepackage{lipsum}
\usepackage{fancyhdr}       %
\usepackage{graphicx}       %
\graphicspath{{media/}}     %

\title{A Relevance Model for Threat-Centric Ranking of Cybersecurity Vulnerabilities
\thanks{\textit{This work was unfunded and performed as part of McCoy's Ph.D. dissertation work at Old Dominion University.}}
}

\author{
  Corren McCoy \\ \\
   \\
  G2 Ops \\
  Virginia Beach, VA 23452\\
  \texttt{corren.mccoy@g2-ops.com} \\
   \And
  Ross Gore \\ \\
  Office of Enterprise Research and Innovation \\
  Old Dominion University \\
  Suffolk, VA 23435\\
 \texttt{rgore@odu.edu} \\
  \And
  Michael L. Nelson, Michele C. Weigle\\ \\
  Dept. of Computer Science \\
  Old Dominion University \\
  Norfolk, VA \\
  \texttt{\{mln, mweigle\}@cs.odu.edu} \\
}

\begin{document}
\maketitle

\begin{abstract}
The relentless process of tracking and remediating vulnerabilities is a top concern for cybersecurity professionals. The key challenge is trying to identify a remediation scheme specific to in-house, organizational objectives. Without a strategy, the result is a patchwork of fixes applied to a tide of vulnerabilities, any one of which could be the point of failure in an otherwise formidable defense. Given that few vulnerabilities are a focus of real-world attacks, a practical remediation strategy is to identify vulnerabilities likely to be exploited and focus efforts towards remediating those vulnerabilities first. The goal of this research is to demonstrate that aggregating and synthesizing readily accessible, public data sources to provide personalized, automated recommendations for organizations to prioritize their vulnerability management strategy will offer significant improvements over using the Common Vulnerability Scoring System (CVSS). We provide a framework for vulnerability management specifically focused on mitigating threats using adversary criteria derived from MITRE ATT\&CK. We test our approach by identifying vulnerabilities in software associated with six universities and four government facilities. Ranking policy performance is measured using the Normalized Discounted Cumulative Gain (nDCG). Our results show an average 71.5\% - 91.3\% improvement towards the identification of vulnerabilities likely to be targeted and exploited by cyber threat actors. The return on investment (ROI) of patching using our policies results in a savings of 23.3\% - 25.5\% in annualized costs. Our results demonstrate the efficacy of creating knowledge graphs to link large data sets to facilitate semantic queries and create data-driven, flexible ranking policies. 
\end{abstract}


\section{Introduction}
\label{sec:introduction}
The relentless process of tracking and prioritizing vulnerabilities for patching is a top concern for cybersecurity professionals \cite{sheyner2002automated}. Ideally every organization would apply the security updates for their operating systems and critical applications as soon as possible after updates are released. However, since patches from top vendors are delivered in monthly blocks on Patch Tuesday, system administrators often find it difficult to select which patches to apply and identify which ones are not applicable \cite{krebs2020, coble2020}. Patch Tuesday is the term used to refer to second Tuesday of each month when Microsoft, Adobe, Oracle, and others regularly release software patches for their software products \cite{wiki:Patch_Tuesday}. Vulnerability prioritization is further hampered when companies delay the automatic installation of security updates in case the patch proves more troublesome than expected \cite{morphisec, threatpost2021}. 

Successful vulnerability management must balance two opposing goals: (1) coverage (fix everything that matters), and (2) efficiency (delay or deprioritize what does not matter) \cite{foreman2010}. In industry, the most prevalent vulnerability management strategy identifies the base CVSS scores for all identified vulnerabilities and patches them in descending score order (10 highest to 0 lowest) \cite{first-epss, first-epss-percentiles, mell2007complete}. Unfortunately, research has shown that CVSS scores are not strongly linked to the emergence of new cyber exploits and system administrators can be overwhelmed by the volume of vulnerabilities with nearly indistinguishable high scores \cite{allodi2014comparing}. While a CVSS score is indicative of vulnerability severity, it does not predict the exploit potential of the underlying software flaw or the operational impact to the organization.

The goal of this research is to demonstrate that aggregating and synthesizing readily accessible, public data sources, can provide an automated patch priority ranking by understanding what vulnerabilities and adversaries are relevant to an organization. Furthermore we show that this proposed strategy offers significant improvements over using  CVSS base scores. 

Our proposed relevance-based ranking model enables businesses to adopt a proactive strategy for vulnerability management \cite{mccoy2022relevance}. Such an approach delivers the most efficient use of people, tools, time, and dollars to address cyber threats that pose the greatest operational risk. Just as search engines provide a better ranking of results based on personalization, so will the ranking of vulnerabilities. Within this context, we seek to define an approach to cybersecurity vulnerability mitigation that improves upon rankings that employ strategies based on the global CVSS metrics associated with known software vulnerabilities published in the National Vulnerability Database (NVD) \cite{NVD}. 

In Section \ref{sec:data_and_methods}, we describe the public data sources we aggregate, the methods we use to synthesize them, and our framework for ranking software vulnerabilities with respect to different organizations for patching. In Section \ref{sec:eval_and_results} we evaluate our approach and present the results. Section \ref{sec:discussion} discusses how our contributions are positioned in the landscape of  software vulnerability management research and identifies several limitations to our work.  Ultimately, we conclude our study in Section \ref{sec:conclusions}.

\section{Data and Methods}
\label{sec:data_and_methods}
Our goal is to remediate vulnerabilities in the most efficient way possible. This requires leveraging, associating, and analyzing different sources of cyber threat intelligence.  The relationships among them need to be understood and they must be organized into a structure for analysis that supports generating prioritized recommendations for effective vulnerability management. 

\begin{figure}[h!]
	\begin{center}
		\includegraphics[width=0.9\textwidth]{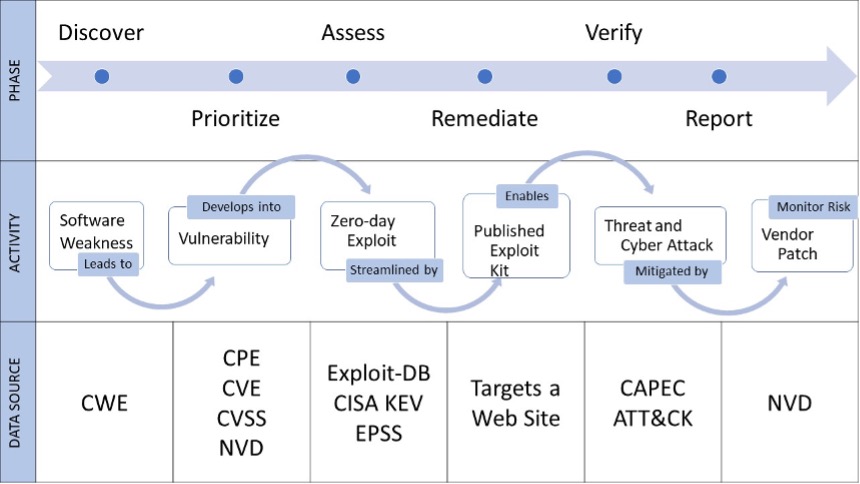}
		\caption{Software vulnerability lifecycle phases and their relationships to our public data sources.}
		\label{fig:vulnerability-model}
	\end{center}
\end{figure}

These data sources are used to model software vulnerabilities with respect to the skill level of a cyber adversary and their motivation to target a specific industry domain (e.g., national defense, higher education, finance, health care). The relationships among these data sources and the software vulnerability lifecycle are summarized in Figure \ref{fig:vulnerability-model}. Specifically, these data sets are:

\begin{enumerate}
    \item The Common Weakness Enumeration (CWE) captures data related to the discovery of a software weakness.
    \item  Data from the Common Vulnerabilities and Exposures (CVE), and Common Vulnerability Scoring System (CVSS) prioritize a vulnerability's severity.
    \item The Exploit Database (ExploitDB), Department of Homeland Security’s Cybersecurity and Infrastructure Security Agency's Known Exploited Vulnerabilities (KEV) catalog, and Exploit Prediction Scoring System (EPSS) assess the likelihood of a software vulnerability being exploited in the wild.
    \item The Common Attack Pattern Enumeration and Classification (CAPEC), and MITRE ATT\&CK knowledge base provides data on how to remediate and mitigate published exploits. 
    \item The National Vulnerability Database (NVD) catalogs and reports vendor provided patches to vulnerabilities in commercial or open source software.
\end{enumerate}

Next, we describe these data sources in more detail and the specific attributes from them that we leverage. Then we highlight how we synthesize them together in a knowledge base to connect data about how an adversary capability can exploit a vulnerability to execute a cyber attack on an organization.
\subsection{Data}
\label{subsec:data}
\subsubsection{Software Weaknesses Dataset}
The Software Weaknesses Dataset consists of data from the CWE. The CWE provides a common language for describing security weaknesses in software architecture, design, or code. It is an encyclopedia of hundreds of types of software weaknesses including: buffer overflow, directory traversal, OS injection, race condition, cross-site scripting, hard-coded password, and insecure random numbers. Each weakness has a technical impact of which only eight technical impacts lead to failure.  These are the weaknesses we focus on. The technical impacts of software weaknesses that lead to failure are: (1) read data, (2) modify data, (3-4) deny service: unreliable execution and resource consumption, (5) execute unauthorized code or commands, (6) gain privileges / assume identity, (7) bypass protection mechanism, and (8) hide activities.

\subsubsection{Vulnerability Dataset}
The Vulnerability Dataset is based on linking entries in the CVE with scoring information from the CVSS. The CVE is the authoritative source of publicly known vulnerabilities. The CVSS is an international standard for measuring the severity of a vulnerability. The CVSS base score is composed of metrics that reflect the intrinsic characteristics of the vulnerability. Each CVE entry includes a unique identifier (CVE number), a short free-text description, and a list of references to additional details of the vulnerability (in the form of URLs). We include this information in our dataset and link it with the CVSS base scores for the vulnerability. 

\subsubsection{Vendor Product Dataset}
The Vendor Product Dataset is based on the CPE. Each entry (i.e. CPE-ID) defines a specific hardware device, operating system, or application software. We excluded entries marked as deprecated and restricted the CPE-IDs of interest to those written in US English. Our Vendor Product Dataset contains more than 15,000 CPE entries representing more than 3,000 products from approximately 200 vendors.

\subsubsection{Attack Pattern Dataset}
Our Attack Patterns Dataset includes 545 unique instances of CAPEC identifiers. CAPEC is a comprehensive dictionary and classification taxonomy of known attacks that can be used by analysts, developers, testers, and educators to advance community understanding and enhance defense sponsored by the United States Department of Homeland Security. A CAPEC identifier can be linked to the MITRE ATT\&CK Enterprise tactics, techniques, and sub-techniques. ATT\&CK provides a common taxonomy for both offense and defense, and has become a standard across many cyber security disciplines to convey threat intelligence, perform testing through red teaming or adversary emulation, and improve network and system defenses against intrusions.

\subsubsection{Exploit Dataset}
Our Exploit dataset is based on a one-to-many mapping between an identified exploit kit (ExploitDB) to the vulnerabilities that are the target of that exploit (CVE). ExploitDB is updated daily and provided by MITRE. It is augmented with the data from CISA's Known Exploited Vulnerabilities (KEV) and the EPSS. The CISA KEV provides real-time updates via email alerts when a newly, identified CVE-ID is exploited. The EPSS model is based on observations of  exploitation attempts against
vulnerabilities, analysis of ancillary information about each of those vulnerabilities, then uses historical events to make predictions about future ones. The EPSS score associated with a CVE-ID represents the probability [0-1] of exploitation in the wild in the next 30 days (following score publication) and the percentile of the current score compared to all scored vulnerabilities with the same or lower EPSS score.

\subsubsection{Adversary Tactics and Techniques Dataset}
The combination of MITRE ATT\&CK and CAPEC datasets form our adversary Tactics and Techniques Dataset. The MITRE ATT\&CK matrices are focused on network defense and describe the operational phases in an adversary's lifecycle. The matrices also detail the specific tactics, techniques, and procedures (TTPs) that Advanced Persistent Threat (APT) groups use to execute their objectives while targeting, compromising, and operating inside a network. Attack patterns enumerated by CAPEC are employed by adversaries through specific techniques described by MITRE ATT\&CK.  Our dataset is formed by linking the CAPEC attack patterns and the related MITRE ATT\&CK techniques together. This enables contextual understanding of the attack patterns within an adversary's operational lifecycle. 

\begin{figure}
	\begin{center}
		\includegraphics[width=0.8\textwidth]{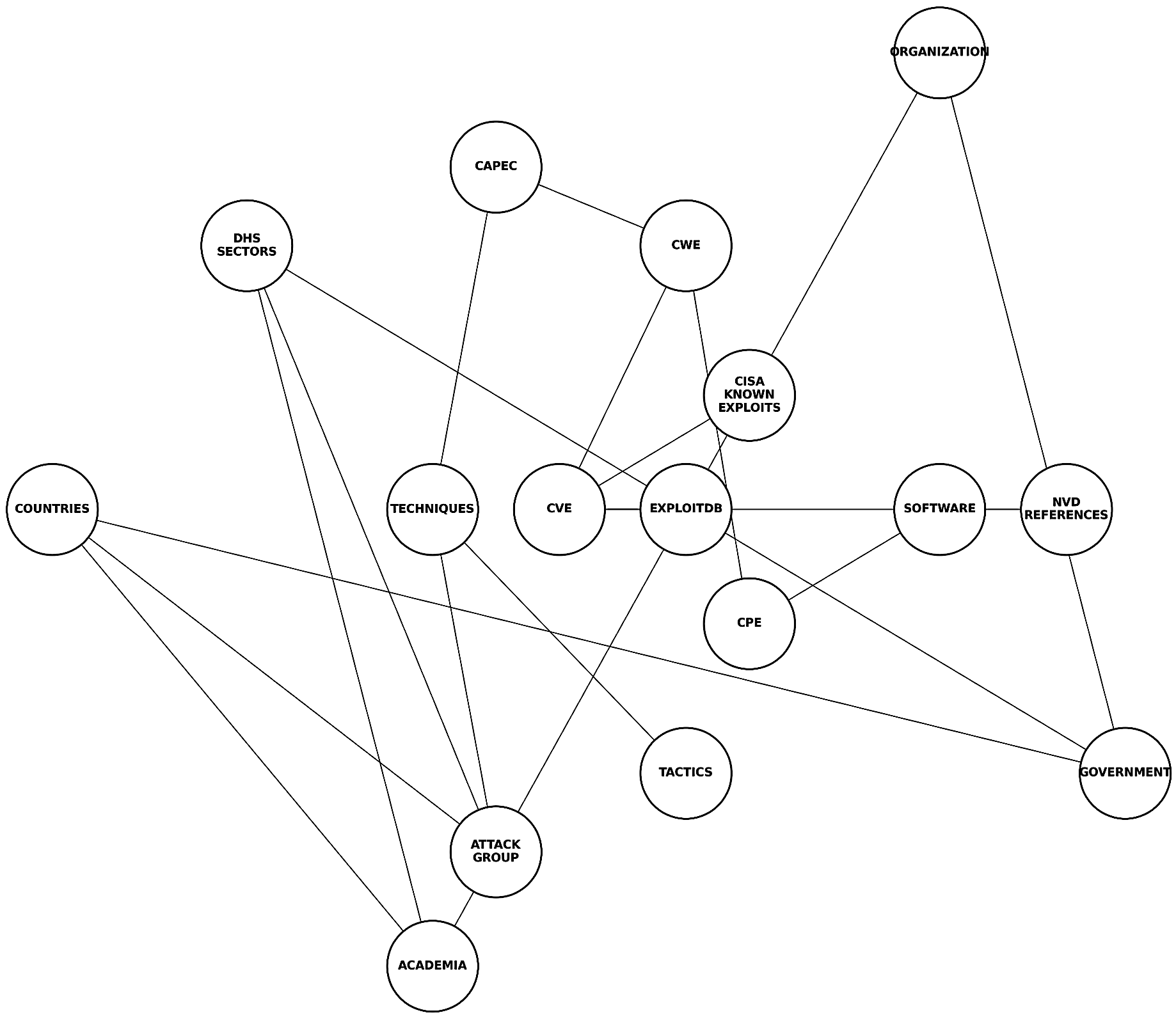}
		\label{fig:graphSchemaEntities}
	\end{center}
 	\caption{The entities of the knowledge graph.}
\end{figure}

\begin{figure}
	\begin{center}
		\includegraphics[width=0.9\textwidth]{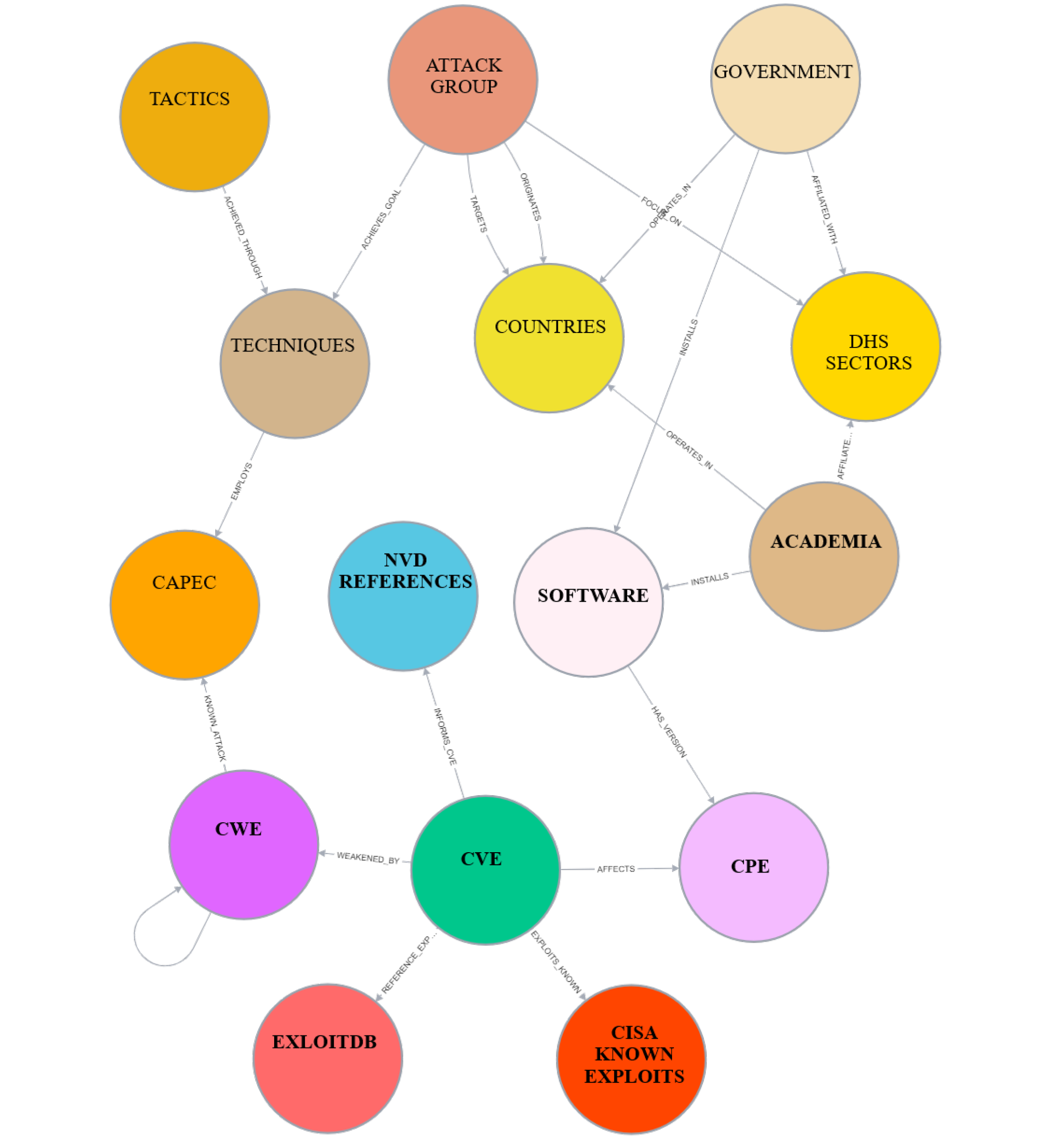}
		\caption{Graph schema representing the type of entities via color in the knowledge graph and the relationship between them.}
		\label{fig:graphSchema}
	\end{center}
\end{figure}

\subsection{Synthesizing Data Sources Into A Knowledge Graph}
\label{sec:synthesize}
The datasets described in Subsection \ref{subsec:data} can be combined to form a knowledge graph. The purpose of the graph is to support queries to effectively rank vulnerabilities for mitigation. This organizational structure is needed. A wealth of information about what vulnerabilities are targeted, who exploits those vulnerabilities, and how they do so currently exists. However, this information is not organized into a structure that comprehensively defines the relationships among the datasets. Our knowledge graph addresses this deficiency. Its overall schema is described in Figure \ref{fig:graphSchema} and Table \ref{tab:schema-rel-color}.

\begin{table}[H]
\caption{Legend for node labels and relationships in knowledge graph schema.}
\begin{tabularx}{1.0\textwidth}{XXX}
	\label{tab:schema-rel-color}\\
	\toprule
	\textbf{Label}       & \textbf{Relationship} & \textbf{Label}    \\
	\midrule
	NVD CVE        & REFERENCE EXPLOIT & ExploitDB                  \\ 	*\midrule
	NVD CVE            & EXPLOITS KNOWN    & CISA Exploit Catalog      \\ 	*\midrule
	NVD CVE          & WEAKENED BY       & CWE                         \\ *\midrule
	CWE           & KNOWN ATTACK      & CAPEC                     \\ *\midrule
	CAPEC        & EMPLOYS            & Attack Enterprise Techniques \\ *\midrule
	Attack Groups    & ACHIEVES GOAL        & Attack Enterprise Techniques  \\ *\midrule
	Attack Groups   & ORIGINATES         & Countries              \\ *\midrule
	Attack Groups  & TARGETS            & Countries                  \\ *\midrule
	Attack Groups & FOCUS ON          & DHS Sectors               \\ *\midrule
	Attack Enterprise Tactics  & ACHIEVED THROUGH     & Attack Enterprise Techniques    \\ *\midrule
	NVD CVE        & AFFECTS            & CPE                        \\ *\midrule
	DHS Sectors    & AFFILIATED WITH   & Organizations             \\ *\midrule
	Organizations & OPERATES IN       & Countries               \\ *\midrule
	Organizations & INSTALLS           & Software              \\ *\midrule
	Software      & HAS VERSION       & CPE                       \\ *\midrule
	NVD CVE       & INFORMS          & NVD References             \\  *\midrule
\end{tabularx}
\end{table}

\subsection{Leveraging the Knowledge Graph to Link Vulnerabilities to Sector Specific Threat Actors}
Our knowledge graph enables us to link vulnerabilities to APTs targeting sectors within the United States and outside of the United States. Here we describe what data processing is required to populate the knowledge graph, and how once populated it links the data together. 

\subsubsection{Defining a Standard Set of Sectors}
The critical infrastructure (CI) sectors denoted by the Department of Homeland Security (DHS) reflect assets, systems, and networks that are vital enough to the United States that incapacitating or destroying them would have a debilitating effect on national security, economics, public health or public safety \cite{cisa-sectors}. Sectors can also be divided into subsectors \cite{dhs-subsectors}.  The CI sectors and subsectors are used to provide an affiliation for both threat actors and the organizations they target in the knowledge graph.

\subsubsection{Defining Standard Locations}
Our knowledge graph requires a standard nomenclature to determine the country or region of origin for cyber threat actors and country of residence for organizations they target for attack. To meet this requirement we leverage the U.S. State Department list of independent states. In this list, the term ``independent state'' refers to people politically organized into a sovereign state with a definite territory recognized as independent by the U.S. \cite{independent-states}. 

\subsubsection{Assigning Attributes to Adversary Groups}
APTs are an extension of nation-states' military forces because of the potential damages and chaos caused by successful critical infrastructure cyber attacks. MITRE keeps track of the APTs. Currently it lists 129 threat groups \cite{mitre-attack2022} in their \emph{Enterprise Framework} that can be associated with known techniques. Using their defined threat profiles, we identify adversaries or threat groups who employ the same tactics and techniques.

\subsubsection{Where Attacks Originate}
For each APT group description provided by MITRE, we use natural language processing to extract keywords to determine the country or independent state from which the group operates. For example, a \emph{North Korean state-sponsored threat group} would be assigned to North Korea with our mapping. We also mined the descriptions to determine year of origin (e.g., 2008) to ascertain the potential longevity of each group. If a year was not explicitly stated in the description, we used the creation date of the MITRE description (e.g., has been active since at least 2009).  

\subsubsection{Who Attacks Each Sector}  
Next, we identify adversarial groups relevant to organizations based on who they target for attacks by mapping APTs and their country to DHS critical infrastructure sectors. To accomplish this we extract the subject of the term `targets', `targeted', or `targeting' in the group description from MITRE. Our knowledge graph includes those where the United States is a targeted country. This choice reflects that our focus is on attacks aimed at the United States. The attribution of APTs to sectors is shown in Table \ref{tab:attack-group-sectors}. Some groups target more than one sector. 

\begin{table}[H]
\caption{DHS sectors ranked by the number of attack groups targeting those sectors based on mentions in MITRE ATT\&CK.}
\begin{tabularx}{1.0\textwidth}{XX}
	\label{tab:attack-group-sectors} \\
	\toprule
		\textbf{Sector Name} & \textbf{Groups Targeting} \\ 
	\midrule
		Government Facilities & 50 \\
		Information Technology & 33 \\
		Financial Services & 19 \\
		Healthcare and Public Health & 17 \\
		Defense Industrial Base & 14 \\
		Energy & 14 \\
		Critical Manufacturing & 10 \\
		Communications & 9 \\
		Transportation Systems & 7 \\
		Chemical & 2 \\
		Water and Wastewater Systems & 1 \\
		Nuclear Reactors, Materials, and Waste & 1 \\
		Emergency Services & 0 \\
		Dams & 0 \\
		Commercial Facilities & 0 \\
		Food and Agriculture & 0 \\ 
	\bottomrule
\end{tabularx}
\end{table}

\subsection{Relevance-Based Ranking Model}
Recall that our goal is to define an approach to cybersecurity vulnerability mitigation that improves upon rankings that employ strategies based on the global CVSS metrics associated with known software vulnerabilities published in the NVD. The outcome is a relevance-based ranking model that can be employed before an adversary takes advantage of a particular vulnerability. The model requires the following components:

\begin{itemize}
	\item Profiles that describe the organization under evaluation in terms of the DHS sector and country in which they operate.
	\item Collection and normalization of a complete software inventory for each organization.
	\item Threat-centric ranking policy definitions based on attack groups of interest and their skill level.
	\item Scoring method for each ranking policy.
\end{itemize}

\subsection{Creating Organizational Profiles}
A vulnerability ranking policy needs to consider the installed software for the organization under evaluation. We identify and define a representative set of organizations in the Government and Education Facilities to serve as organizational benchmarks for evaluating our vulnerability management approach.

\subsection{Software Used in the Education Subsector}
CollegeSimply\cite{collegesimply} provides a list of Virginia colleges and sources public domain college data from the U.S. Department of Education National Center for Education Statistics. Using the list, we chose six universities of varying sizes and funding sources (public and private). The public universities are the University of Virginia (UVA), Virginia Tech (VT), Old Dominion University (ODU), and William \& Mary University (WM), and the private universities are Regent University (REGENT) and Washington and Lee University (WLU). For each university,  we located a published list of supported academic software on the university's web site. Then we assign CPE-IDs to each piece of software. The full academic software listing is provided in \cite{mccoy2022relevance}. A summary of the number of vulnerabilities found in the academic software associated with each university is shown in Table \ref{tab:academicSoftware-cpe}. We assign ``Size'' (small (S), medium (M), large (L),  extra large (XL)) based on the number of software products that were publicly listed. However, this does not reflect the size of the university or the number of software products that the university actually uses.

\begin{table}[H]
\caption{Academic software associated with vendor product CPE-ID.}
\begin{tabularx}{1.0\textwidth}{X X X X X X X }
	\label{tab:academicSoftware-cpe} \\
	\toprule
		& \textbf{WM} & \textbf{ODU}	& \textbf{VT} & \textbf{RU} & \textbf{UVA}	& \textbf{WLU} \\
	\midrule
	CPEs   & 24 	& 47     & 12 	    & 23 	 & 30 	    & 13 \\ 
	Soft.  & 33 	& 69     & 22     	& 31 	 & 49    	& 23 \\ *\midrule
	Size & Medium & Extra Large 	& Small	& Medium	& Large 	& Small \\ 
	\bottomrule
\end{tabularx}
\end{table}

\subsection{Software Used by Government Facilities}
Government facilities do not routinely publish the software they use. However, the National Information Assurance Acquisition Policy, NSTISSP No. 11 \cite{NSTISSP}, requires government agencies to purchase only commercial security products that have met specified third-party assurance requirements and have been tested by an accredited national laboratory\cite{NSTISSP}.\footnote{The list of certified products is available at \url{https://www.commoncriteriaportal.org/products/}.}  In accordance with NSTISSP, the \emph{Common Criteria} is an internationally recognized set of guidelines (ISO 15408) \cite{ISO.15408} that define a common framework for evaluating security features and capabilities of Information Technology security products against functional and assurance requirements. 

We reduced Common Criteria to the set of products certified for use in the U.S. Then we searched for CPE-IDs across all categories based on the vendor and product name. The software list shown in Table \ref{tab:governmentSoftware-cpe}, consisting of applications and operating systems, was generated by randomly selecting software from the Common Criteria with assigned CPE-IDs in groups of 14, 24, 30, and 47 to approximately match the cardinality of the small (S), medium (M), large (L), and extra-large (XL) university software lists. 

\begin{table}[H]
\caption{Government facility software associated with vendor product CPE-ID.}
\begin{tabularx}{1.0\textwidth}{XXXXX}
	\label{tab:governmentSoftware-cpe} \\
	\toprule
	& \textbf{GOV-S} & \textbf{GOV-M} & \textbf{GOV-L} & \textbf{GOV-XL} \\ 
	\midrule
	Software Assigned & 14 & 24 & 30 & 47 \\
	Common Criteria & 57 & 57 & 57 & 57 \\
\bottomrule
\end{tabularx}
\end{table}

\subsection{Ranking Policy Definitions}
Deciding which vulnerabilities to remediate is a daunting task. In a perfect world, all vulnerabilities would be remediated as they were discovered, but unfortunately that does not happen in reality. An exploit observed in the wild is the most relevant proxy for the probability that an exposed vulnerability can be used to compromise an organization's network. To that end the predictive ranking policies we evaluate identify candidate vulnerabilities that fit the pattern of known attack groups. Formally, this is the intersection of vulnerabilities in the software used by an organization and vulnerabilities being actively targeted by threat actors. 

The criteria for our ranking policies using the attacker characteristics and targets is discussed in Subsection \ref{sec:synthesize}. Each policy leverages data points in the knowledge graph to provide a scoring methodology that considers:

\begin{itemize}
	\item Which threat actors use the same technique to initiate an attack?
	\item Given an industry, which threat actors target it?
	\item Given a type of attack, which vulnerabilities does it exploit?
	\item At present day, what is the probability of exploit?
	\item Given an organization, which vulnerabilities are present in the installed software?
\end{itemize}

We created four different ranking policies that answer these questions. Each policy prioritizes different information based on organizational information preferences with regard to specific threats. The policies also include knowledge on whether an exploit for the CVE-ID has been observed.

\begin{itemize}
	\item \textbf{Policy One: CVSS Base Score Ranking}: Vulnerabilities are remediated based on the assigned CVSS Base Score ranking from most severe (``critical'') to least severe (``low'').  
	\item \textbf{Policy Two: APT Threat Ranking}: Vulnerabilities are remediated based on the likelihood of present day exploit and the existence of a technique employed by an attack group that targets the industry in the country where the organization operates.
	\item \textbf{Policy Three: Generalized Threat Ranking}: Vulnerabilities are remediated based on the likelihood of exploit by a low-skilled or highly-skilled adversary that have high impact on the organization.
	\item \textbf{Policy Four: Ideal Ranking}: The ideal ranking employs the same criteria as the APT and Generalized threat rankings, Policy Two and Three, but has the foreknowledge that a vulnerability has already been exploited using information from the ExploitDB and CISA KEV databases. 
\end{itemize}

\subsection{Ranking Policy Implementations}
\label{sec:policies}
For each CVE-ID, we examine 16 features using the cyber intelligence data sources. The features inform each policy and create a set of relevance scores for ranking CVE-IDs as they are published. The features are: (1) CVE-ID, (2) CVSS Base Score Metrics, (3) Publication date, (4) Modification date, (5) CAPEC-ID, (6) CAPEC skill level, (7) ATT\&CK technique name, (8) MITRE ATT\&CK group id, (9) MITRE ATT\&CK group country of operation, (10) Risk Appetite, (11) EPSS Probability of Exploit, (12) CISA Known Exploit Catalog, (13) Exploit-DB, (14) Organization identifier, (15) Critical infrastructure sector, and (16) Organization's country of residence. The source code used in our implementation of the ranking policies is available in \cite{corren_mccoy_2024_10569299}.

Based on the policy definitions, the CVSS V3.1 Base Score is the only feature needed to implement Policy One. The features needed to implement Policy Two and its ideal version in Policy Four are listed in Table \ref{tab:relevanceAlgo1}. 

\begin{table}[H]
	\caption{Policy Two scoring features using MITRE ATT\&CK data feed to characterize the threat to the organization.}
\begin{tabularx}{1.0\textwidth}{XXX}
	\label{tab:relevanceAlgo1} \\
	\toprule
	\textbf{Feature}	& \textbf{Specific Threat Relevance Rank Value} & \textbf{Ideal Rank Value} \\ 
	\midrule
	CVSS Base Metric (Attack Vector) & Network & Network \\ *\midrule
	DHS Sector & Govt. Facilities, Education &Govt. Facilities, Education \\ *\midrule
	Org Country & United States & United States \\ *\midrule
	Attack Group Country & China, Russia, Iran & China, Russia, Iran \\ *\midrule
	Risk Appetite & [0, 100] &  [0, 100] \\ *\midrule
	EPSS Probability & 0.876 &  \\ *\midrule
	CISA KEV or Exploit-DB  &  &  \\
        Entry Exists & & True \\ *\midrule
	Software Affected & True & True \\ *\midrule
	Scoring Range & [1-6] & [1-6] \\ 
	\bottomrule
\end{tabularx}
\end{table}

For Policies Two, Three, and Four, we use a binary weighting [0,1] for each feature to determine its existence as applicable to a specific CVE-ID. The sum of the categorical values is presented as the relevance score we used to rank the associated CVE-IDs using the logic shown in the algorithm provided in  \cite{mccoy2022relevance}. The minimum assigned relevance score is set to 1 using this algorithm to avoid a long tail of non-relevant CVE-IDs and ensure only relevant CVE-IDs associated with the organization's installed software are candidates for ranking.

When determining what to patch, one must consider both the setup and business disruption costs. Then one has to weigh them against the potential exploitation cost, and decide when and how often to patch an enterprise system or application. The total  costs  of  a  vulnerability are the sum  of  its  direct  costs  (level of  effort employed  by human  resources)  and  indirect  costs  (productivity losses,  interruption  of  production  processes after patching). Previous research has established that these costs can be measured in non-monetary units based on the severity of the vulnerability where Low=0.25, Medium=1, High=1.5, and Critical=3 units  \cite{fruhwirth2009improving}. We evaluate the economic cost of remediating vulnerabilities using these established units.

\section{Evaluation and Results}
\label{sec:eval_and_results}
\subsection{Candidate Generation} 
In our study, we use 55,939 CVE-IDs published between 2019-2021 as the corpus from which to identify a much smaller subset of candidate vulnerabilities for ranking. We use the CVE modification date to simulate examining the vulnerabilities as they were published. A total of 3,079 unique CVE-IDs were found to be applicable across all of the Government Facilities (Table \ref{tab:total-vulns-govt}) and Education subsector (Table \ref{tab:total-vulns-education}) software lists. The data and source code used in our evaluation is available in \cite{corren_mccoy_2024_10569299}.

For the Government Facilities shown in Table \ref{tab:total-vulns-govt}, we observed low, annual vulnerability counts for three of the four proxy organizations at less than 2\% of all CVE-IDs analyzed. Even the largest government organization, GOV-XL, which was designed to mirror the breadth of software (i.e., 47 products) of its counterpart ODU in the Education subsector, experienced less than 4\% of all CVE-IDs analyzed. The low number of vulnerabilities in the sector may be attributed to the selection process for software products assigned to Government Facilities in this study which were selected exclusively from the certified product list approved by the \emph{Common Criteria} \cite{NSTISSP}. This outcome may provide an indication that the rigor imposed upon these products in terms of security requirements and on-going evaluation may potentially reduce their exposure to published vulnerabilities.

\begin{table}[H]
\caption{Total vulnerabilities by year for government facilities sector.}
\begin{tabularx}{1.0\textwidth}{lrrrr}
	\label{tab:total-vulns-govt}\\
	\toprule
	\textbf{Year}  & \textbf{GOV-S} & \textbf{GOV-M} & \textbf{GOV-L} & \textbf{GOV-XL} \\ 
	\midrule
	2019  & 8              & 41             & 51             & 102             \\
	2020  & 11             & 34             & 55             & 144             \\
	2021  & 16             & 84             & 140            & 285             \\ * \midrule
	Total Vulns & 35    & 159   & 246   & 531    \\
	Pct of All Vulns & 0.25\%    & 1.15\%   & 1.77\%   & 3.85\%    \\
\bottomrule
\end{tabularx}
\end{table}

\begin{table}[H]
\caption{Weekly vulnerability traffic by year for government facilities sector.}
\begin{tabularx}{1.0\textwidth}{@{}llrrrr@{}}

	\label{tab:weekly-vuln-govt}\\
	\toprule
	\textbf{Year} &
	\textbf{Org} &
	\textbf{Avg Vuln Per Week} &
	\textbf{Min Vuln Per Week} &
	\textbf{Max VulnPer Week} &
	\textbf{Weeks Per Year} \\
        \midrule
	2019 & GOV-XL & 4 & 1 & 20 & 32 \\
	2019 & GOV-L  & 3 & 1 & 9  & 24 \\
	2019 & GOV-M  & 2 & 1 & 4  & 23 \\
	2019 & GOV-S  & 2 & 1 & 2  & 5  \\  * \midrule
	2020 & GOV-XL & 4 & 1 & 13 & 40 \\
	2020 & GOV-L  & 3 & 1 & 10 & 23 \\
	2020 & GOV-M  & 3 & 1 & 10 & 16 \\
	2020 & GOV-S  & 2 & 1 & 3  & 6  \\ * \midrule
	2021 & GOV-XL & 7 & 1 & 25 & 43 \\
	2021 & GOV-L  & 4 & 1 & 14 & 40 \\
	2021 & GOV-M  & 3 & 1 & 11 & 29 \\
	2021 & GOV-S  & 2 & 1 & 3  & 10
	\\
        \bottomrule
\end{tabularx}
\end{table}

For the Education subsector shown in Table {\ref{tab:total-vulns-education}, we also observed low vulnerability counts of less than 2\% for organizations with small amounts of reported software, such as VT and WLU. %
Conversely, we note that universities who reported more software in use such as ODU, REGENT, and WM need to evaluate hundreds of vulnerabilities as candidates for remediation during the course of any given year.

\begin{table}[H]
\caption{Total vulnerabilities by year for education subsector.}
\begin{tabularx}{1.0\textwidth}{@{}rrrrrrr@{}}

	\label{tab:total-vulns-education}\\
	\toprule
	\multicolumn{1}{r}{\textbf{Year}} &
	\multicolumn{1}{r}{\textbf{VT}} &
	\multicolumn{1}{r}{\textbf{WLU}} &
	\multicolumn{1}{r}{\textbf{REGENT}} &
	\multicolumn{1}{r}{\textbf{WM}} &
	\multicolumn{1}{r}{\textbf{UVA}} &
	\multicolumn{1}{r}{\textbf{ODU}} \\ * \midrule
		2019           & 14 & 3 & 1396  & 1370 & 188   & 1457 \\
	2020           & 6  & 57  & 565  & 556  & 279  & 751  \\
	2021           & 15 & 144 & 1721  & 1704 & 639 & 2026 \\* \midrule
	Total Vulns    & 35 & 204 & 3682 & 3630 & 1106 & 4234 \\* 
	Pct of All Vulns & 0.25\% & 1.45\% & 26.56\% & 26.19\% & 7.98\% & 30.54\% \\
        \bottomrule
\end{tabularx}
\end{table}

\begin{table}[H]
\caption{Weekly vulnerability traffic by year for education subsector.}
\begin{tabularx}{1\textwidth}{@{}llrrrr@{}}
	\label{tab:weekly-vuln-edu}\\
	\toprule
	\textbf{Year} &
	\textbf{Org} &
	\textbf{Avg Vuln Per Week} &
	\textbf{Min Vuln Per Week} &
	\textbf{Max Vuln Per Week} &
	\textbf{WeeksPer Year} \\* \midrule
 	2019 & VT     & 2  & 1 & 4   & 7  \\
  	2019 & WLU    & 1  & 1 & 1   & 3  \\
   	2019 & REGENT & 35 & 1 & 442 & 40 \\
	2019 & WM     & 43 & 1 & 441 & 32 \\
	2019 & UVA    & 11 & 1 & 44  & 18 \\
	2019 & ODU    & 35 & 1 & 444 & 42 
 \\ \midrule
 	2020 & VT     & 1  & 1 & 1   & 6  \\
  	2020 & WLU    & 5  & 1 & 20  & 12 \\
   	2020 & REGENT & 15 & 1 & 57  & 40 \\
	2020 & WM     & 15 & 1 & 58  & 39 \\
	2020 & UVA    & 9  & 1 & 34  & 33 \\
	2020 & ODU    & 18 & 1 & 59  & 43 
 \\ \midrule
	2021 & VT     & 3  & 1 & 4   & 7  \\
 	2021 & WLU    & 7  & 1 & 23  & 21 \\
	2021 & REGENT & 36 & 1 & 264 & 48 \\
	2021 & WM     & 36 & 1 & 258 & 48 \\
	2021 & UVA    & 15 & 1 & 120 & 43 \\
        2021 & ODU    & 41 & 1 & 315 & 50 \\* \bottomrule
\end{tabularx}
\end{table}

Figures \ref{fig:vuln-month-gov} shows the accumulated vulnerabilities by month and year for each organization in our study. It is important to note the unpredictable manner in which newly published and modified CVE-IDs can present themselves for analysis and remediation to an organization.

\begin{figure}
	{\label{gov-eval-2019}\includegraphics[width=0.5\linewidth]{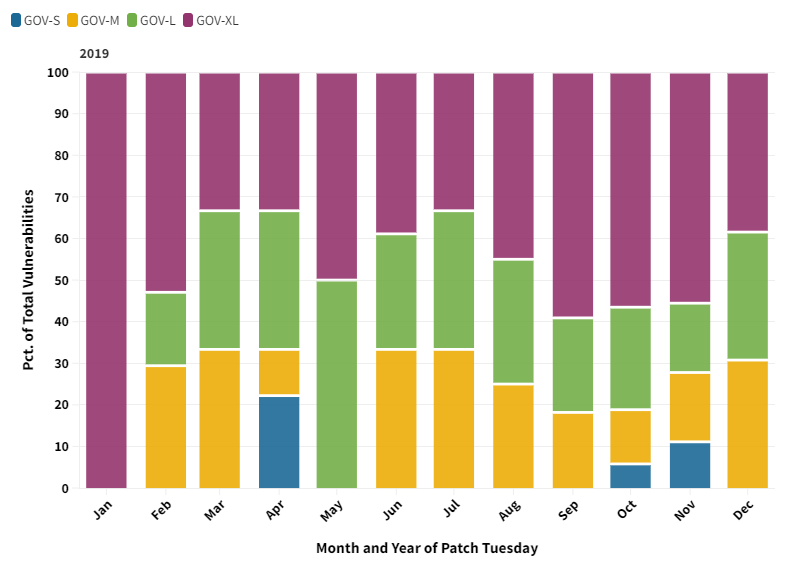}}	
        {\label{edu-eval-2019}\includegraphics[width=0.5\linewidth]{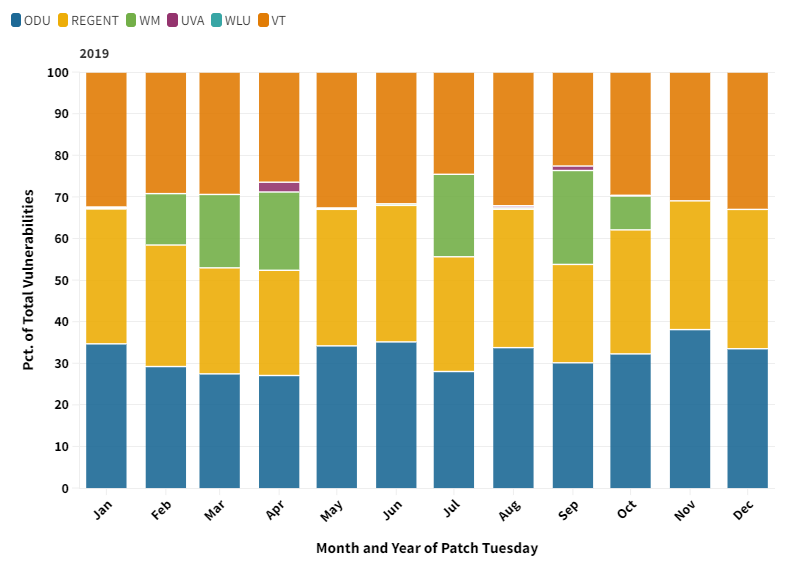}}
        {\label{gov-eval-2020}\includegraphics[width=0.5\linewidth]{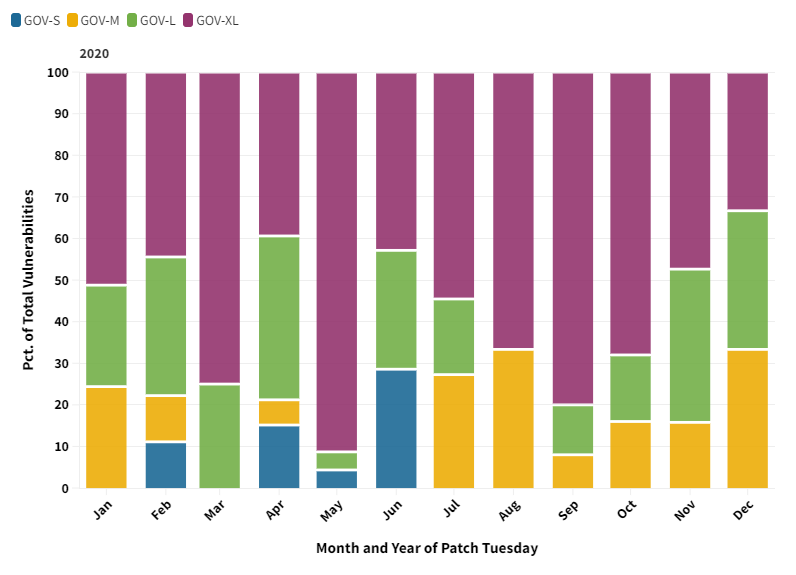}}
         {\label{edu-eval-2020}\includegraphics[width=0.5\linewidth]{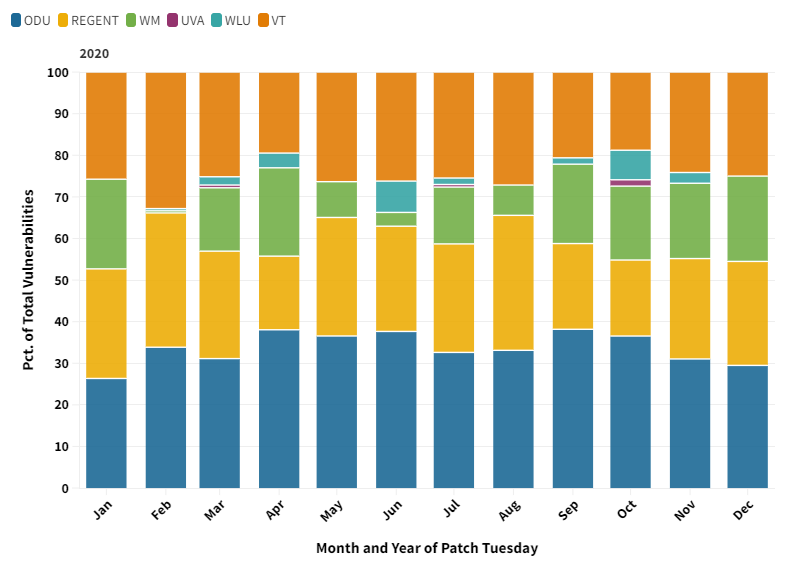}}
        {\label{gov-eval-2021}\includegraphics[width=0.5\linewidth]{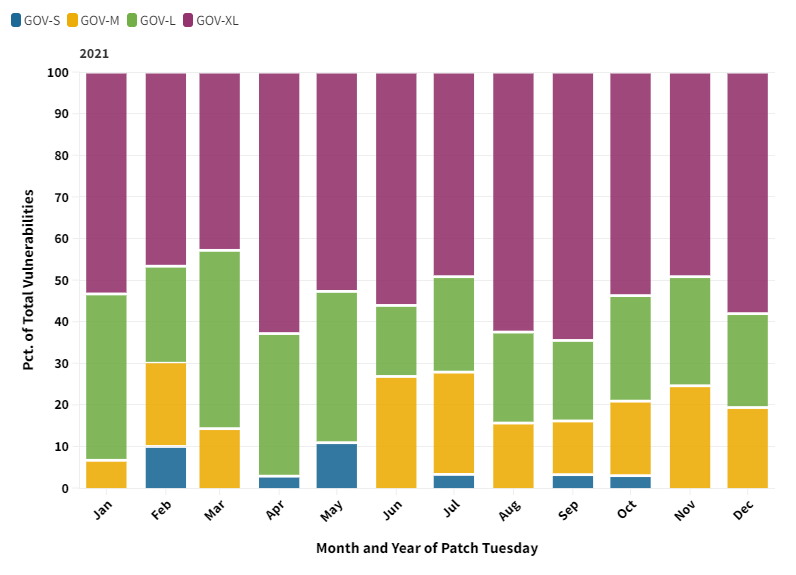}}
        {\label{edu-eval-2021}\includegraphics[width=0.5\linewidth]{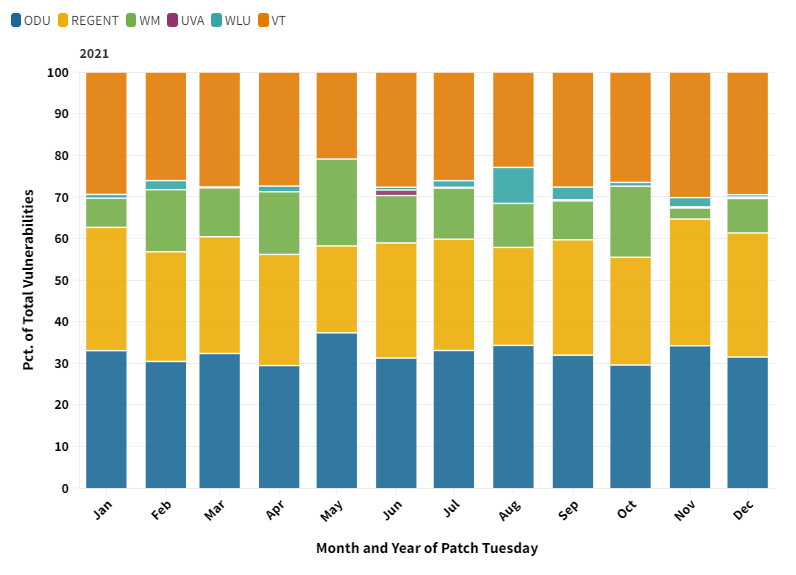}}
	\label{fig:vuln-month-gov}
 \caption{Vulnerabilities by month and year for CVE-IDs between 2019 and 2021 for government facilities (left) and education (right) sectors.}
\end{figure}

\subsection{Normalized Discounted Cumulative Gain}

Within the field of cybersecurity there is no consensus approach for measuring, testing, and comparing the accuracy of a ranking model. Therefore, this research, similar to others in discussed in Section \ref{sec:discussion}, builds upon measurements derived from the field of Information Retrieval (IR). Evaluation measures for IR assess how well the search results from a recommender satisfy a given query. Specifically, recommender systems use the Normalized Discounted Cumulative Gain (nDCG) \cite{jarvelin2002cumulated} score to evaluate the ranking of items (e.g., individual vulnerabilities) in a collection (e.g., NVD). 

The nDCG varies from 0.0 to 1.0, with 1.0 representing the ideal ranking order. The nDCG is commonly used to evaluate Search Engine Result Pages (SERPs) where the position of an entry indicates its search result relevancy. Higher ranked pages are more likely to gain the attention of the consumer. We apply the same approach towards creating a ranking list for patching vulnerabilities. Order is important to ensure higher ranked CVE-IDs are considered first.  The main difficulty encountered when using nDCG is the availability of an ideal ordering of results when feedback (e.g., human judgment) is not available. We address this shortcoming faced by SERPS with Policy Four, introduced in Section \ref{sec:policies} as a data-driven proxy of an ideal ordering of vulnerabilities. 

In order to compare the results of rankings between each relevance policy and the ideal ranking (Policy Four), we calculated the nDCG of each CVE-ID for every organizational interaction with our ranking system. The nDCG values were averaged for each weekly collection of CVE-IDs to obtain a measure of the average performance of our ranking algorithms.  The application of nDCG in this study is interpreted as follows:

\begin{enumerate}
	\item ``G'' is for gain -- it corresponds to the magnitude of each vulnerability's relevance.
	\item ``C'' is for cumulative -- it refers to the cumulative gain, or summed total, of every vulnerability's relevance score.
	\item ``D'' is for discounted -- it divides each vulnerability's scored relevance by the scored relevance of the associated ideal policy to reflect our goal of having the most relevant vulnerabilities ranked towards the top of our mitigation lists.
	\item ``N'' is for normalized -- it divides Discounted Cumulative Gain (DCG) scores by ideal DCG scores calculated for a ground truth data set, as represented by the relevance scores and ranking resulting from our ideal policy (i.e. Policy Four), which  used foreknowledge of exploited vulnerabilities contained within historical ExploitDB and CISA KEV intrusion detection reports. 
\end{enumerate}

Once the relevance value is computed for each CVE-ID, we rank each entry based on the relevance value and compute the nDCG using the formulas shown below.
\begin{equation}
DCG_k=\sum_{i=1}^k\frac{2^{rel_i}-1}{\log_2{\left(i+1\right)}}
\end{equation}

The cumulative gain at K is the sum of gains of the first K items recommended. \(iDCG_k\) is the maximum possible (ideal) \(DCG\) for a given set of queries, vulnerabilities, and relevance scores.

\begin{equation}
nDCG_k=\frac{DCG_k}{iDCG_k}
\end{equation}

The chart in Figure \ref{fig:ndcg-k_analysis} illustrates the average values of nDCG for each position K based on weekly vulnerability collections. Recall, K reflects the number of CVE-IDs to remediate. The number of observations ranges from 383 when K=1 to 16 when K=100. The x-axis reports the rank (from 1 to 100), while the y-axis displays the respective value of nDCG@K. Figure \ref{fig:ndcg-k_analysis} shows that the CVSS Base Score performs moderately well at the ends of the spectrum when K=1 and K=100. However, the performance decreases when 5 $\leq$ K $\leq$ 50. Policy Two (APT Threat) is not impacted by the number of weekly CVE-IDs; it performs at a consistent level regardless of the number of CVE-IDs encountered.

\begin{figure}[H]
	\begin{center}
		\includegraphics[width=1.0\textwidth]{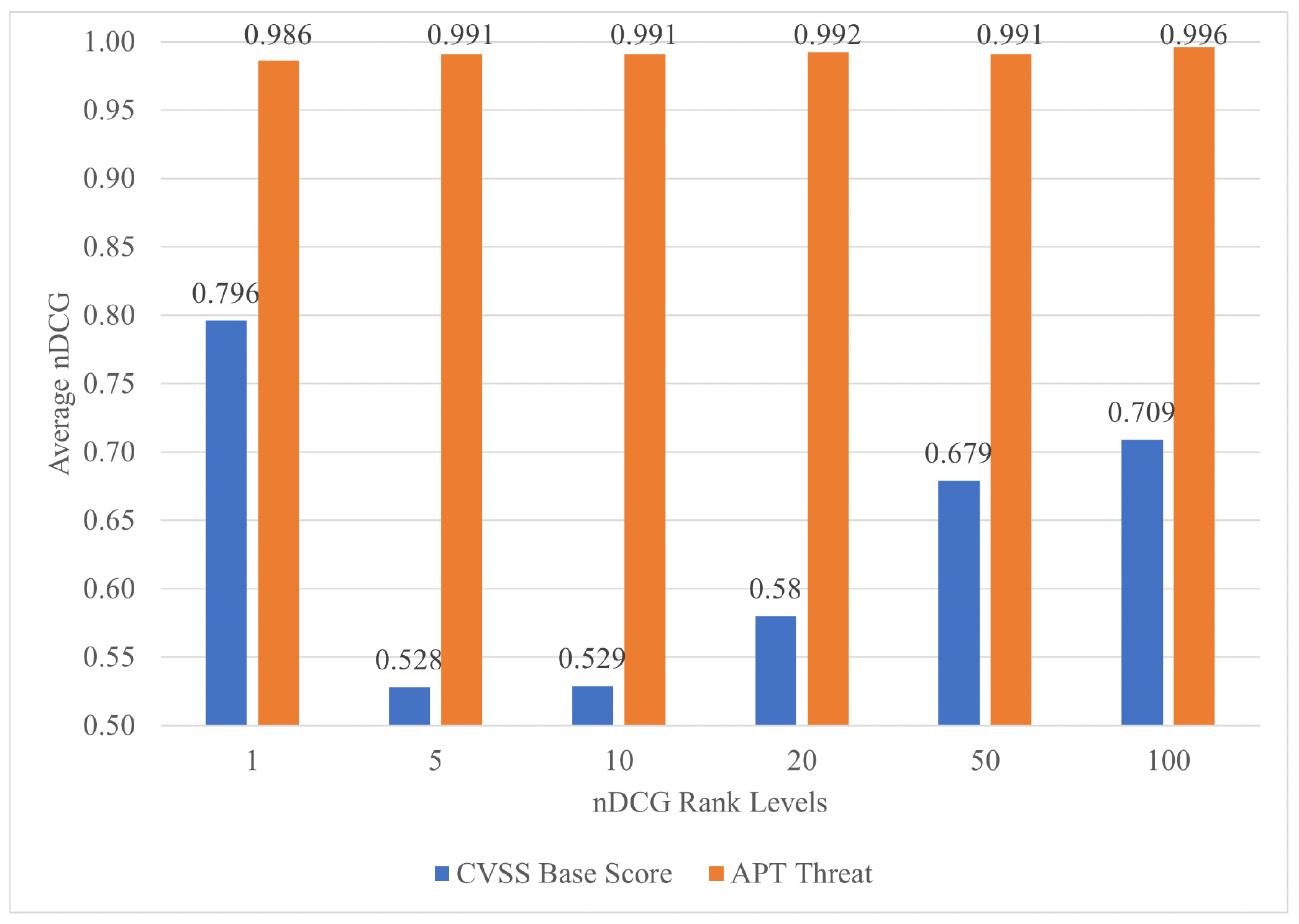}
		\caption{Average value of nDCG at different rank levels (K) for CVSS Base Score versus APT Threat policy for the ODU, REGENT, and WM organizations.}
		\label{fig:ndcg-k_analysis}
	\end{center}
\end{figure}

\begin{table}[H]
	\caption{Average performance of Policy One (CVSS Base Score) versus Policy Two (APT Threat) where China is the source region of interest (nDCG@20).}
\begin{tabularx}{1\textwidth}{@{}lcccc@{}}

	\label{tab:ranking-eval-china}\\
	\toprule
	\textbf{Year} &
	\multicolumn{1}{c}{\textbf{CVSS Base Score}} &
	\multicolumn{1}{c}{\textbf{APT Threat China}} &
	\multicolumn{1}{c}{\textbf{Avg. Difference in nDCG}} &
	\multicolumn{1}{c}{\textbf{Known Exploits}} \\* \midrule
	\multicolumn{5}{c}{\textbf{ODU}}     \\* \midrule
	2019 & 0.601 & 0.996 & 0.394 & 4 \\
	2020 & 0.557 & 0.998 & 0.441 & 2  \\
	2021 & 0.571 & 0.986 & 0.415 & 12 \\* \midrule
	\multicolumn{5}{c}{\textbf{REGENT}}  \\* \midrule
	2019 & 0.592 & 0.999 & 0.407 & 2  \\
	2020 & 0.557 & 0.998 & 0.441 & 1      \\
	2021 & 0.585 & 0.985 & 0.399  & 12    \\* \midrule
	\multicolumn{5}{c}{\textbf{WM}}    \\* \midrule
	2019 & 0.598 & 0.998 & 0.400  & 3     \\
	2020 & 0.565 & 0.998 & 0.433 & 1    \\
	2021 & 0.585 & 0.985 & 0.399  & 12  \\* \bottomrule
\end{tabularx}
\end{table}

In our results, the nDCG@20 measures for Policy One, the CVSS Base Score policy, were in the range of [0.343, 1], as shown in Figure \ref{fig:ndcg-weekly-vulns-apt}.  Lower values for nDCG@20 were observed with Policy One when the number of vulnerabilities collected exceeded the minimum threshold (i.e. 20) by more than 1000\% (e.g., 200+). Higher nDCG@20 values were observed when the number of vulnerabilities was closer to the threshold (e.g., 20 to 30). Policy Two, the APT threat-centric policy, was minimally impacted by the number of vulnerabilities and ranged from [0.878, 1].

\begin{figure}[H]
\caption{nDCG@20 for Policy One (CVSS Base Score) versus Policy Two (APT Threat) for the ODU, REGENT, and WM organizations.}
	\begin{center}
		\includegraphics[width=0.8\textwidth]{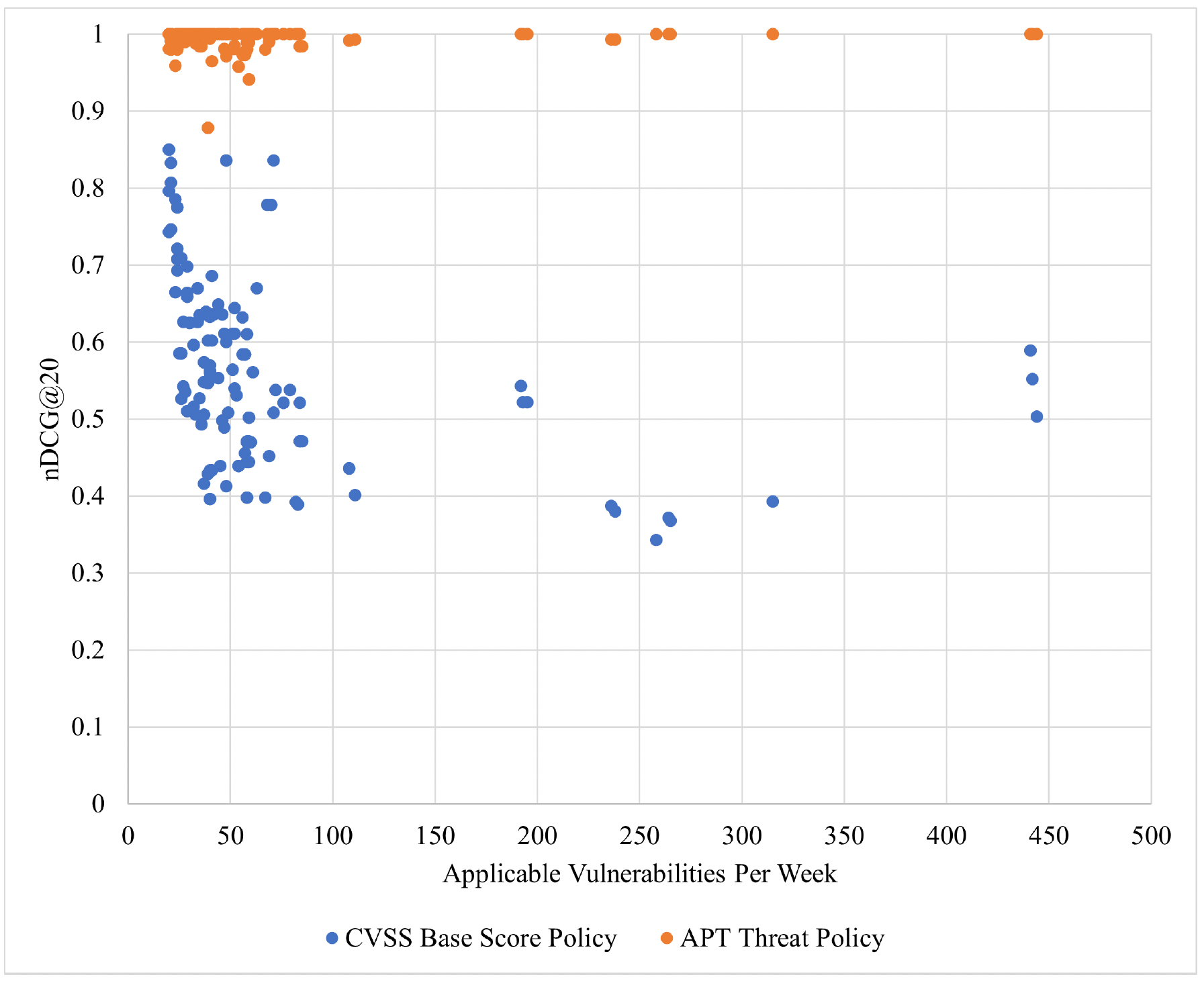}
		\label{fig:ndcg-weekly-vulns-apt}
	\end{center}
\end{figure}

Table \ref{tab:ranking-eval-capec} shows similar results for Policy Three, our general threat-centric policy. The average difference in nDCG@20 of 0.35 indicates Policy Three performs 91.3\% better than Policy One, the CVSS Base Score policy, as an indicator of vulnerabilities that might be targeted by a highly-skilled cyber threat actor. 

\begin{table}[H]
	\caption{Average performance of Policy One (CVSS Base Score) versus Policy Three (General Threat) with a highly skilled adversary (nDCG@20).}
\begin{tabularx}{1.0\textwidth}{@{}lcccc@{}}

	\label{tab:ranking-eval-capec}\\
	\toprule
	\textbf{Year} &
	\textbf{CVSS Base Score} &
	\textbf{General Threat Highly Skilled} &
	\textbf{Avg. Difference in nDCG} &
	\textbf{Known Exploits} \\* \midrule
	\bottomrule
	\multicolumn{5}{c}{\textbf{ODU}}           \\* \midrule
	\textbf{2019} & 0.543 & 0.988 & 0.444 & 4  \\
	\textbf{2020} & 0.548 & 0.998 & 0.450  & 2  \\
	\textbf{2021} & 0.474 & 0.986 & 0.511 & 12 \\* \midrule
	\multicolumn{5}{c}{\textbf{REGENT}}        \\* \midrule
	\textbf{2019} & 0.528 & 0.995 & 0.467 & 2  \\
	\textbf{2020} & 0.512 & 0.999 & 0.487 & 1  \\
	\textbf{2021} & 0.500 & 0.984 & 0.484 & 12 \\* \midrule
	\multicolumn{5}{c}{\textbf{WM}}            \\* \midrule
	\textbf{2019} & 0.538 & 0.992 & 0.454 & 3  \\
	\textbf{2020} & 0.520 & 0.999 & 0.478 & 1  \\
	\textbf{2021} & 0.499 & 0.984 & 0.485 & 12 \\* \bottomrule
\end{tabularx}
\end{table}

\begin{figure}[H]
	\begin{center}
		\includegraphics[width=0.8\textwidth]{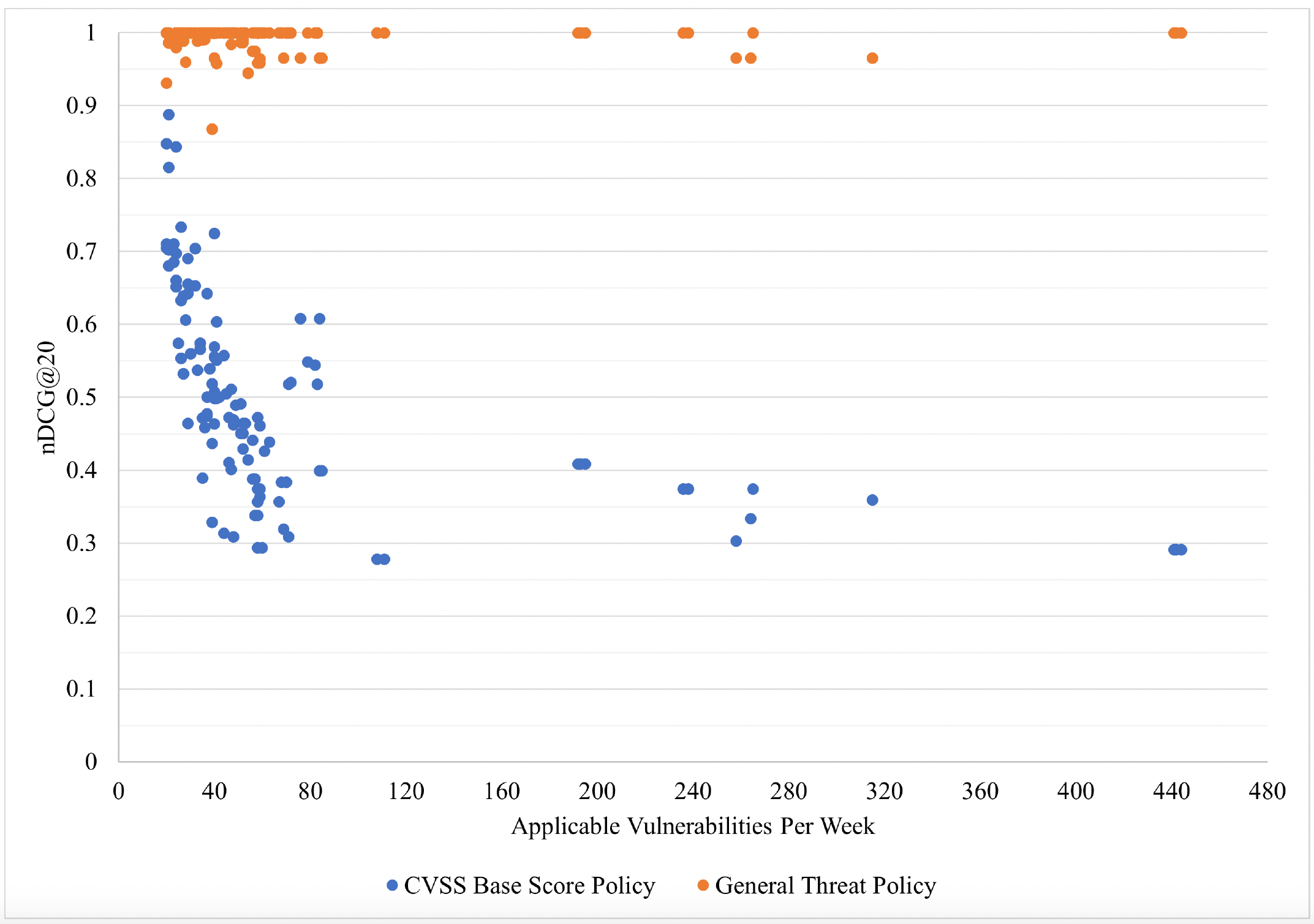}
		\caption{nDCG@20 for the Policy One (CVSS Base Score) versus Policy Three (General Threat) for the ODU, REGENT, and WM organizations.}
		\label{fig:ndcg-weekly-vulns-general}
	\end{center}
\end{figure}

Using all of the weekly observations (n=163) across organizations, we perform a paired t-test \cite{wiki:paired_student} to compare the mean of the nDCG for Policy One (CVSS Base Score) against Policy Two (APT Threat). Results of the paired t-test indicate that there is a significant large difference between Policy One (CVSS Base Score) [Mean = 0.58, STDEV = 0.1] and Policy Two (APT Threat) [Mean = 0.992, STDEV = 0.02] and the p-value equals 0. The Policy Two (APT Threat) population's nDCG@20 average is greater than the Policy One (CVSS Base Score) population's average, and the difference is large enough to be statistically significant. 

We perform a similar test to compare the mean of the nDCG for Policy One (CVSS Base Score) against Policy Three (General Threat). Results of the paired t-test indicate that there is a significant large difference between Policy One (CVSS Base Score) [Mean = 0.512, STDEV = 0.139] and Policy Three (General Threat) [Mean = 0.99, STDEV = 0.022] and the p-value equals 0. The Policy Three (General Threat) population's average nDCG@20 greater than the Policy One (CVSS Base Score) population's average, and the difference is large enough to be statistically significant.

These results show that CVSS base score metrics do not contain a data element or scoring component that allows for enumeration of a specific threat. The paired t-test indicates that the difference in the recommended ranking positions of CVE-IDs between policies is statistically significant (p-value equals 0). Therefore, any relevance ranking based solely on the CVSS base score will fall short of the organization's specified goals. These results also provide another indication that the severity of a vulnerability, as measured by its CVSS base score, may not be the optimal ranking approach for every organization.

\subsection{Cost of Patch Prioritization}
Past research has shown that organizations cannot fix all of their known vulnerabilities. Instead, they are able to fix 5\% - 20\%  \cite{first-epss} of known vulnerabilities per month. Here, we examine the annualized cost of remediating the top 20 vulnerabilities produced by our different ranking policies: Policy One (CVSS Base Score), Policy Two (APT threat-centric), and Policy Three (General Threat Centric). Recall, we use the non-monetary units, defined by Fhurwith et al., associated with patching. The results of this analysis are shown in Table \ref{tab:cost-eval-china}.  In all cases, there is decreased average cost of 23.3\% when Policy Two (APT threat-centric) is used for prioritizing CVE-IDs for remediation. Specifically, Policy Two realizes (APT threat-centric) decreases of 498 units for ODU, 390.5 units for REGENT, and 455.75 units for WM over the three year evaluation period when compared to Policy One (CVSS Base Score).

\begin{table}[H]
\caption{Difference in the cost of patching the top 20 CVE-IDs for Policy One (CVSS Base Score) versus Policy Two (APT Threat) where China is the source region of interest.}
\begin{tabularx}{1.0\textwidth}{@{}lccc@{}}

	\label{tab:cost-eval-china}\\
	\toprule
	\textbf{Year} &
	\textbf{CVSS Base Score Cost} &
	\textbf{APT Threat China Cost} &
	\textbf{Average Savings} \\* \midrule
	\multicolumn{4}{c}{\textbf{ODU}}           \\* \midrule
	\textbf{2019} & 631.50  & 449.25 & 182.25 \\
	\textbf{2020} & 531.00  & 439.00    & 92.00  \\
	\textbf{2021} & 994.50  & 770.00    & 224.50 \\* \midrule
	\multicolumn{4}{c}{\textbf{REGENT}}        \\* \midrule
	\textbf{2019} & 604.75 & 422.25 & 182.50 \\
	\textbf{2020} & 375.50  & 308.50  & 67.00  \\
	\textbf{2021} & 960.00   & 752.00    & 208.00 \\* \midrule
	\multicolumn{4}{c}{\textbf{WM}}            \\* \midrule
	\textbf{2019} & 603.75 & 424.75 & 179.00 \\
	\textbf{2020} & 374.00    & 308.50  & 65.50  \\
	\textbf{2021} & 960.00    & 748.75 & 211.25 \\* \bottomrule
\end{tabularx}
\end{table}

Table \ref{tab:cost-eval-capec} shows increased savings in patch costs using Policy Three (General Threat). The cost of patching remains the same across all organizations using the CVSS Base Score. However, for each organization Table \ref{tab:cost-eval-china} shows additional savings over using Policy Two (APT Threat). Specifically, decreases of 548.25 units for ODU, 500.75 units for REGENT, and 499.75 for WM represent an average 25.6\% improvement over the CVSS Base Score approach. Recall, Policy Two (APT Threat) only provided a 23.3\% improvement.

\begin{table}[H]
	\caption{Difference in the cost of patching the top 20 CVE-IDs for Policy One (CVSS Base Score) versus Policy Three (General Threat) from a highly skilled adversary.}
\begin{tabularx}{1\textwidth}{@{}lccc@{}}
	\label{tab:cost-eval-capec}\\
	\toprule
	\textbf{Year} &
	\textbf{CVSS Base Score Cost} &
	\textbf{General Threat Cost} &
	\textbf{Average Savings} \\* \midrule
	\multicolumn{4}{c}{\textbf{ODU}}           \\* \midrule
	\textbf{2019} & 631.50 & 438.50 & 193.00 \\
	\textbf{2020} & 531.00 & 424.50 & 106.50 \\
	\textbf{2021} & 994.50 & 745.75 & 248.75 \\* \midrule
	\multicolumn{4}{c}{\textbf{REGENT}}        \\* \midrule
	\textbf{2019} & 604.75 & 412.00 & 192.75 \\
	\textbf{2020} & 375.50 & 294.50 & 81.00  \\
	\textbf{2021} & 960.00 & 733.00 & 227.00 \\* \midrule
	\multicolumn{4}{c}{\textbf{WM}}            \\* \midrule
	\textbf{2019} & 603.75 & 412.50 & 191.25 \\
	\textbf{2020} & 374.00 & 296.50 & 77.50  \\
	\textbf{2021} & 960.00 & 729.00 & 231.00 \\* \bottomrule
\end{tabularx}
\end{table}

Using all of the weekly observations (n=163) across organizations, we perform a paired t-test \cite{wiki:paired_student} to compare the mean of the patch costs for Policy One (CVSS Base Score) against Policy Two (APT Threat). Results of the paired t-test indicate that there is a significant large difference between Policy One (CVSS Base Score) [Mean = 37.025, STDEV = 10.291] and Poicy Two (APT Threat) [Mean = 28.362, STDEV = 5.475] and the p-value equals 7.45e-27. The population of Policy Two's (APT Threat) average patch cost is considered to be less than Policy One's (CVSS Base Score), and the difference is large enough to be statistically significant. 

Similarly, we perform a paired t-test \cite{wiki:paired_student} to compare the mean of the patch costs for Policy One (CVSS Base Score) against Policy Three (General Threat). Results of the paired t-test indicated that there is a significant large difference between Policy One (CVSS Base Score) [Mean = 37.025, STDEV = 10.291] and Policy Three (General Threat) [Mean = 27.523, STDEV = 4.905] and the p-value equals 9.989e-30. The population of Policy Three's (General Threat) average patch cost is considered to be less than Policy One's (CVSS Base Score), and the difference is large enough to be statistically significant. 

\subsection{Predicting Exploits}
Only a small subset (2\%-7\%) of published vulnerabilities are exploited in the wild \cite{first-epss}. Given that such a small number of CVE-IDs are actually exploited, it is advantageous for organizations to leverage as much insight as possible to identify potential threats. Here, we demonstrate how Policy Two (APT Threat) can be used to prioritize a vulnerability with a known exploit. The ODU organization identified 39 CVE-IDs to mitigate during the week of November 23, 2021. 

In this case study, the top 20 are ranked in accordance with Policy Two (APT Threat) as shown in Table \ref{tab:ODU-TOP20-Example}. We note that three CVE-IDs in this group, CVE-2021-38000, CVE-2021-30632, and CVE-2021-30633, have known exploits. The CISA Known Exploits entry for CVE-2021-38000, which impacts Google Chrome, is shown in Figure \ref{fig:CISA_CVE-2021-38000}. The entries in Table \ref{tab:ODU-TOP20-Example} show that all three CVE-IDs are identified as relevant using Policy Two. However, CVE-2021-38000 is ranked at position 29 using Policy One based on its CVSS base score of 6.1 (medium severity). This highlights that when using Policy One (CVSS Base Score), CVE-2021-38000 falls outside of the top 20 range for remediation by IT administrators at ODU. In contrast, Policy Two (APT Threat) elevates this CVE-ID to position number 3 because of its high relevance score.

\begin{table}[H]
	\caption{Application of ranking policies by ODU for vulnerabilities published during the week of 23-November-2021. Known exploits are bolded and highlighted in grey.}
\begin{tabularx}{1.0\textwidth}{@{}lcccc@{}}
	\label{tab:ODU-TOP20-Example}\\
	\toprule
	\textbf{CVE-ID} &
	\textbf{CVSS Base Score} &
	\textbf{Relevance Score} &
	\textbf{Policy 1 Rank} &
	\textbf{Policy 2 Rank} \\* \midrule
	\bottomrule
	CVE-2021-37966          & 4.3          & 6          & 34          & 1   \\
	CVE-2021-37999          & 6.1          & 6          & 28          & 2 \\
	\textbf{CVE-2021-38000} & \textbf{6.1} & \textbf{6} & \textbf{29} & \textbf{3}  \\
	CVE-2021-30542          & 8.8          & 2          & 5           & 4      \\
	CVE-2021-30543          & 8.8          & 2          & 6           & 5     \\
	CVE-2021-30626          & 8.8          & 2          & 7           & 6     \\
	CVE-2021-30627          & 8.8          & 2          & 8           & 7    \\
	CVE-2021-30628          & 8.8          & 2          & 9           & 8    \\
	CVE-2021-30629          & 8.8          & 2          & 10          & 9      \\
	CVE-2021-30630          & 4.3          & 2          & 31          & 10     \\
	\textbf{CVE-2021-30632} & \textbf{8.8} & \textbf{2} & \textbf{11} & \textbf{11}  \\
	\textbf{CVE-2021-30633} & \textbf{9.6} & \textbf{2} & \textbf{2}  & \textbf{12} \\
	CVE-2021-34423          & 9.8          & 2          & 1           & 13      \\
	CVE-2021-34424          & 7.5          & 2          & 26          & 14        \\
	CVE-2021-37956          & 8.8          & 2          & 12          & 15        \\
	CVE-2021-37957          & 8.8          & 2          & 13          & 16       \\
	CVE-2021-37958          & 5.4          & 2          & 30          & 17       \\
	CVE-2021-37959          & 8.8          & 2          & 14          & 18       \\
	CVE-2021-37961          & 8.8          & 2          & 15          & 19        \\
	CVE-2021-37962          & 8.8          & 2          & 16          & 20     \\* \bottomrule
\end{tabularx}
\end{table}

\begin{figure}[]
	\begin{center}
		\includegraphics[width=0.9\textwidth]{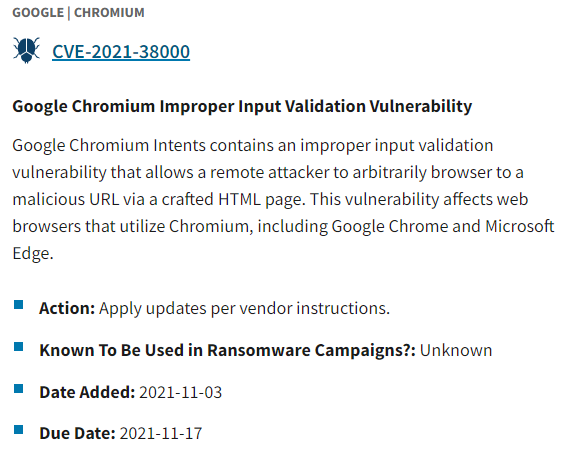}
				\caption{CISA Known Exploits catalog entry for CVE-2021-38000 (Reproduced from \cite{cisa-kev-catalog}).}
		\label{fig:CISA_CVE-2021-38000}
	\end{center}
\end{figure}

\section{Discussion}
\label{sec:discussion}
There is a myriad of existing research that falls within the scope of our work. Here, we discuss related research, identify the limitations of our work and highlight the contributions it provides.

\subsection{Related Research}
Multiple researchers have created ontologies to represent the cyber security domain by aggregating multiple sources of information \cite{hemberg2020bron, bridges2013automatic, jones2015towards, iannacone2015developing, bizer2011linked, wang2009ovm}. This work provides the foundation for building automated tools which reduce the scope, complexity, and volume of security data that must be managed by security professionals which we leverage in our approach. However, our research differs from these efforts in that we extract more information and sources to achieve completeness in our knowledge graph. In addition, categorization is a necessary pre-cursor to our ranking policies for vulnerability management. Multiple research efforts have shown that identifying and categorizing additional metadata about vulnerabilities, exploits, attacks, and targets can be beneficial \cite{fruhwirth2009improving, frei2006large, alberts2003managing, Tsukerman2021}. More recently, the application of text-mining to extra additional data about these entities has led to models which predict the severity of a vulnerability using only text-based data \cite{jacobs2020improving, chen2016xgboost, almukaynizi2017proactive, khazaei2016automatic, OSVDB}.

Even with an organized understanding of the cyber threat domain, understanding how to minimize the cost of managing and protecting information assets is a challenge. A core component of this challenge is adopting a vulnerability management process that can detect and remediate known vulnerabilities \cite{allodi2014comparing}. A common approach is to remediate all vulnerabilities above a certain severity score. However, this approach has been found to be sub-optimal \cite{dey2015optimal} and in some cases, no better than randomly choosing vulnerabilities to remediate \cite{allodi2014comparing}. Furthermore, in many cases it is infeasible to patch all the CVEs with the highest CVSS base scores due to the time and resources required for remediation actions. This is because 13.5\% of the NVD vulnerabilities are scored between 9 and 10 \cite{alperin2019risk}.

This has led to extensive work in evaluating if the CVSS score can be a good predictor for vulnerability exploitation \cite{allodi2012preliminary}, and whether it can be improved by additional information \cite{tatarinova2019extended, notess2002wayback, wayback, horawalavithana2019mentions}. Machine learning approaches have also been explored \cite{jacobs2019exploit, jacobs2020improving} as well as exploit prediction models that leverage data from online sources generated by the white-hat community (i.e., ethical hackers) \cite{almukaynizi2017proactive}. Vulnerability  exploitation  can also be  modeled  as a transition between system states \cite{singhal2017security, ou2006scalable, ou2005mulval, homer2009sound, gallon2011cvss, noel2014big, noel2015big}. However, these graphs often tend to be unwieldy as network size grows, making the identification of realistic paths to compromise difficult to achieve \cite{obes2013attack}. Customized and target specific ranking approaches also exist \cite{alperin2019risk, allodi2017towards, allodi2014comparing, allodi2017work}. However, these approaches assume the existence of site-specific threat intelligence information.

\subsection{Contributions of our approach}
Prior research has demonstrated the ability to examine adversary capabilities, vulnerability management, and exploit prediction at a particular point in time or with isolated threat scenarios. However, little research has been done to create an end-to-end prioritization approach which encompasses the entire vulnerability management life cycle. We address this gap by:

\begin{itemize}
	\item Extracting dozens of essential features about the vulnerability, including its potential for harm, the degree to which it is exploitable, and how frequently the vulnerability is targeted by adversaries.
	\item Leveraging the ability of property graphs to offer a flexible schema where one can add attributes to extend the data model, create hierarchies with different levels of granularity, and combine multiple dimensions to better manage big data. 
	\item Performing an assessment of current and predicted future attacker activity based on known tactics and techniques.
	\item Correlating threat and exploit intelligence from publicly available authoritative sources.
	\item Devising an approach to convert raw data about threat indicators into contextual risk scores.
	\item Identifying how important the affected asset is to an organization in any industry.
	\item Inferring indirect facts and hidden relationships which can further inform our results.
\end{itemize}

Parsing real-time, open source cyber threat intelligence data cannot be accomplished by a human analyst. Therefore, we automate its correlation and analysis using a knowledge graph. We can also leverage the Application Programming Interfaces (APIs) and data feeds maintained by NIST to provide awareness of the changing threat landscape while allowing for dynamic and continuous assessment of the underlying network architecture. Our research provides benefits to organizations seeking to create high-level strategies to examine cybersecurity posture in a manner that is predictive not just reactive.

\subsection{Known Limitations}
It is important to note that our work is not without limitations. To apply our approach, organizations must have a methodology to accurately construct a software inventory that can be correlated with an entry in the Common Platform Enumeration (CPE) database. Vulnerabilities cannot be allocated without a CPE-ID and low fidelity inventory reporting may result in residual cyber risk. The relevance ranking policies we have identified can only be effectively applied to a known software architecture. Furthermore, it is important to note that the attack group list in MITRE ATT\&CK is not all encompassing. A Google search will identify emerging APT groups that are not included in the MITRE's enterprise matrices. In addition, the proof of concept code entries collected via ExploitDB does not include a time component which indicates when the POC entry was made. As a result, it is not possible to discretely link the CVE-ID's publication or modification date with the subsequent appearance of an intrusion report. The inclusion of a timestamp would have allowed us to evaluate the predictive portion of our policies based on a timeline of events. Our approach is naive with regard to exploitation and does not consider the publication date for exploit code maturity using ExploitDB. The ExploitDB to CVE mapping webpage is also not well covered in the Internet Archive. 

Time lapse dynamics related to data sources exist, too. The EPSS probability scores and percentiles are dynamic and should be collected near the time of the CVE publication date. In order to maintain consistency in our data set, all cyber intelligence data was collected and frozen for analysis as of 31-December-2021. Future work can utilize the API provided by the EPSS team to dynamically collect the scores and percentiles in real time. This is a candidate for future work. Finally, the optimum ordering approach that we prescribe may not ease patch hesitancy or prevent a culture of ``wait and see'' with regard to patching vulnerabilities. Our policies also cannot control the quality of vendor patch distributions on Patch Tuesday that in some cases lead to recalls later in the month. These scenarios are outside the scope of this research. Our ranking policies can, however, reduce the amount of unnecessary work spent patching CVE-IDs that are neither applicable nor associated with a known cyber threat actor.

\section{Conclusions}
\label{sec:conclusions}
The process of tracking and remediating vulnerabilities is relentless. The key challenge is trying to identify a remediation scheme specific to in-house, organizational objectives. Without a strategy, the result is a patchwork of fixes applied to a tide of vulnerabilities, any one of which could be the point of failure in an otherwise formidable defense. The goal of this research is to demonstrate that aggregating and synthesizing readily accessible, public data sources to provide personalized, automated recommendations for organizations to prioritize their vulnerability management strategy will offer significant improvements over the current state of the art solutions. Our results showed an average 71.5\% - 91.3\% improvement towards the identification of vulnerabilities likely to be targeted and exploited by cyber threat actors. The ROI of patching using our policies results in a savings in the range of 23.3\% - 25.5\% in annualized costs. A paired t-test demonstrates these findings are statistically significant and offer an improvement over the industry standard approach to vulnerability management. 

Overall, the relevance ranking strategy described in this study emphasizes the capability of threat-centric scenarios for ranking and prioritizing vulnerabilities with due consideration of the threat environment. We have shown that a network defender, who typically has to address thousands of exposed vulnerabilities, can spend fewer resources to patch more vulnerabilities that are much more likely to be exploited and of interest to a specific set of cyber threat actors. The automated data aggregation within our knowledge graph allows one to submit queries to identify new vulnerabilities that affect the most important software and servers. This ability to differentiate among vulnerabilities and how they might be targeted by an adversary enhances the state of the art in vulnerability management.

\section*{Acknowledgments}
This work was unfunded and performed as part of McCoy's Ph.D. dissertation work at Old Dominion University.

\end{document}